\documentclass[11pt,twocolumn]{article}

\usepackage[dvips]{graphicx}
\usepackage{natbib}
\begin{document}

%
%
\title{Conditional Multilevel Monte Carlo Simulation of Groundwater Flow in the Culebra Dolomite at the Waste Isolation Pilot Plant (WIPP) Site}

\author{M.~Park \and
		K.A.~Cliffe 
        }
\author{M. Park\thanks{University of Nottingham (corresponding author, min.park@nottingham.ac.uk)} 
\and K. A. Cliffe\thanks{University of Nottingham. Deceased January 2014.}}
\date{}

\maketitle

%
%
%

%
%
\begin{abstract}
We extended the multilevel Monte of Carlo (MLMC) approach to simulation of groundwater flow in porous media by incorporating direct measurements of medium properties. Numerical simulations of Waste Isolation Pilot Plant (WIPP) repository in southeastern New Mexico are performed to test the performance of the conditional MLMC technique. The log-transmissivity of WIPP site is modeled as the conditional random fields which honor exact field values at a few locations. The conditional random fields are generated through the modified circulant embedding methods in \citep{dietrich:1996}. We also study effects of a combination of the conditional MLMC accompanied by antithetic variates. The main quantity of interest is the time of radionuclides travelling from the center of repository to the site boundary. Numerical examples are presented to demonstrate the cost-effectiveness of the multilevel approach in comparison to the standard Monte Carlo (MC) simulation. 
\end{abstract}

%
%
\section{Introduction}
Understanding and managing groundwater systems are essential in the safety assessment of geological disposal facilities for radioactive waste as leaked radionuclides are transported through groundwater flow. With rapid development of computational capability, computer simulations have become an indispensable modeling tool for the groundwater research. Generally, the simulation of groundwater flow involves intrinsic uncertainties due to the lack of complete knowledge of porous medium at all locations, i.e., only a few measurement of medium properties such as porosity, hydraulic conductivity and transmissivity are available, and values at other locations are subject to uncertainty. As a result, this uncertainty propagates throughout the calculations, and quantification of its impact on results of computer simulation is the core issue of ongoing research in this field.

The commonly used approach to quantifying uncertainty in groundwater flow is to regard the hydraulic conductivity as a random field with given mean and spatial correlation structure \citep{marsily:2005,delhomme:1979}. One of the essential challenges in this approach is how to solve elliptic PDEs with random coefficients efficiently. There have been several attempts to use stochastic Galerkin and stochastic collocation based on polynomial chaos method \citep{xiu:2010,maitre:2010}. However, these methods are usually very expensive when the number of stochastic degrees of freedom required in the model is large, while truncating the number of random variables to any computationally feasible number results in large errors.

The standard Monte Carlo technique is also one of widely used methods for solving stochastic partial differential equations (SPDEs) arising from mathematical modeling of groundwater flow. However, the computational cost of Monte Carlo simulation can be very high for large-scale problems and its rate of convergence is notoriously slow. To overcome these difficulties, \citet{cliffe:2011} and \citet{barth:2011} employed the multilevel Multilevel Monte Carlo (MLMC) method for the uncertainty quantification in the groundwater flow simulation using unconditional log-permeability fields. 	 

The conditional simulation of the flow and transport in heterogeneous porous media widely studied in many works \citep{dagan:1982,graham:1989,gotway:1994,lu:2004}. The primary effect of conditioning on given measurements of hydraulic parameters is reduction of variation, i.e. uncertainty, in a neighborhood of observed data. Thus, in turn, the subsequent statistical error of the results can be more efficiently attenuated by conditioning.   

The purpose of this paper is to study the impact of the use of conditional random fields in multilevel Monte Carlo simulation for groundwater flow which is an extension of the works \citep{cliffe:2011, teckentrup:2013}. In addition to variance reduction by conditioning, antithetic variates, which is one of widely used variance reduction techniques for the standard Monte Carlo methods, is employed for further variance reduction. 

In this paper we study the conditional multilevel Monte Carlo technique through a case study of Waste Isolation Pilot Plant (WIPP) repository in the Culebra Dolomite, New Mexico. Section 2 gives the description of the mathematical model of groundwater flow. In Section 3 contains descriptions of the conditional random field generation, the MLMC technique, and the antithetic variates method. Section 4 represents the results of the WIPP site simulations and investigates the cost-effectiveness of MLMC over standard method.  

\section{Mathematical Model}
The rate of groundwater flow through a porous media is related to the porous medium properties and the gradient of the hydraulic head, which can be written using Darcy{'}s law as
%
\begin{equation} \label{darcy}
q = -k \nabla u,
\end{equation}
 where $q$ is the Darcy flux, $k$ is the hydraulic conductivity of the porous medium, $u$ is the hydraulic head and $\nabla$ is a gradient operator.
 
The next equation for the groundwater model is the conservation of mass 
\begin{equation} \label{conservation}
\nabla \cdot q = 0,
\end{equation}
where $\nabla \cdot$ is the divergence operator with respect to the spatial coordinates.

As mentioned before, in order to quantify uncertainty in $q$ and $u$, the hydraulic conductivity is modeled as a (conditional or unconditional) random field $k(\textbf{x},\omega)$ on $D \times \Omega$ with a certain mean and covariance structure. Here, $D \in \mathrm{R}^d$ is a bounded spatial domain and $\Omega$ is the set of all possible outcomes. Then the governing equation describing the groundwater fluid can be derived by combining Darcy's law (\ref{darcy}) with the mass balance (\ref{conservation}), which can be written as 
\begin{equation} \label{eq:groundwater}
\nabla \cdot (k(\textbf{x},\omega)\nabla u(\textbf{x},\omega) = 0.
\end{equation}

If the thickness of thin layer of rock is relatively small compared to its lateral extent, it is often appropriate to assume that groundwater flow is two dimensional. As a result, the hydraulic conductivity, $k$ , in equation (\ref{eq:groundwater}) can be replaced by transmissivity via $T = k b$, and this yields the equation: 

\begin{equation} \label{eq:2dgroundwater}
\nabla \cdot (T(\textbf{x},\omega)\nabla u(\textbf{x},\omega) = 0 \quad \mbox{in } D \subset \mathbf{R}^2.
\end{equation}
The dimension reduction of three-dimensional groundwater flow equation to two-dimensional results in simulations with significantly smaller computer memory requirements and with shorter computer execution times.

Hence, we consider the problem of solving the equation (\ref{eq:groundwater}) over a domain $D$, given the boundary conditions on the boundary $\partial D$: the Dirichlet condition on a part of $\partial D$, i.e.,
\begin{equation} \label{eq:dirichle} 
u = g_D(\textbf{x}) \mbox{ on } \Gamma_D, 
\end{equation}
and the Neumann condition on the remain part of $\partial D$, 

\begin{equation} \label{eq:neumann}
\frac{\partial u}{\partial \mathbf{n}} = g_N(\textbf{x}) \mbox{ on } \Gamma_N,
\end{equation}
where $\mathbf{n}$ is the normal vector to the boundary, with $\Gamma_D \cup \Gamma_N = \partial D$.

After the pressure is calculated, the travel time of a particle is found using the transport equation
\begin{equation}\label{eq:transport}
\dot{\textbf{x}}(t) = \frac{q(\textbf{x})}{b \phi} = -\frac{T(\textbf{x})}{b \phi} \nabla u(\textbf{x}),
\end{equation}
with the initial condition
\begin{equation} \label{eq:transport_init}
\textbf{x}(0) = \textbf{x}_0,
\end{equation}
where $\textbf{x}(t)$ is the the location of the particle at time $t$, $b$ is the thickness of thin layer of rock and $\phi$ is the porosity of rock. 

\section{Preliminaries}
\subsection{Random Field Generation } \label{sec:condrand}
In equation (\ref{eq:2dgroundwater}), we model hydraulic transmissivity, $T$, as a two-dimensional random field. The transmissivity values are often assumed (see, e.g. \citep{delhomme:1979, cliffe:2011}) to have the lognormal distribution, i.e. $\log_{10} T = Z \sim \mathcal{N}(\mu, R)$, where $\mathcal{N}(\mu, R)$ denotes the multinormal distribution with mean $\mu$ and covariance $R$. We note that this assumption guarantees that T is positive. For the covariance matrix $R$, we take the following isotropic exponential covariance function  
\begin{equation}\label{eq:exp_cov}
R_{ij} \ = C(\mathbf{x}_i,\mathbf{x}_j) := \ \sigma^2 
\mathrm{exp}\left(-\frac{\|\mathbf{x}_i - \mathbf{x}_j\|_2}{\lambda}\right), 
\end{equation}
where $\mathbf{x}_i,\mathbf{x}_j \in D.$ The parameters $\sigma^2$ and $\lambda$ denote the variance and the correlation length, respectively.

One possible approach to generate a stationary Gaussian random field, $Z$, is to use the matrix decomposition of the covariance matrix $R$ , such as Cholesky decomposition. Although the matrix decomposition method does generate random fields with the exact covariance structure, its computational cost is very high even with a few hundreds of the sampling points in each coordinate direction in two dimensional space. Moreover, round-off error become more significant in a large-scale problem because the covariance matrix is likely to become extremely ill-conditioned \citep{dietrich:1989}. 

The turning band method \citep{gotway:1994} and the Karhunen-Lo$\grave{\mbox{e}}$ve (KL) decomposition method \citep{ghanem:1991} have been widely used for the groundwater applications. However, these two approaches introduce errors due to using a finite number of lines and eigenfunctions respectively, consequently there are an inevitable trade-off between an accuracy and computational cost in both methods.    

The circulant embedding algorithm \citep{dietrich:1997}, on the other hand, is the exact and fast simulation of stationary Gaussian random fields on a rectangular sampling grid, and this is the main reason why we choose the circulant embedding method as an unconditional random field generator in our numerical simulations. The algorithm uses periodic embeddings resulting in the (block) circulant matrices and square roots of this type of matrices can be efficiently constructed via the fast Fourier transform (FFT) method. Each realization of the random fields can then be efficiently generated by multiplying complex-valued Gaussian random vector by this square root. This matrix-vector product can also be rapidly computed using FFT. The only significant limitations of this method are that it is only applicable to the stationary fields on a rectangular grid and that the circulant matrix must be positive definite. To achieve its positive definiteness, one might need to increase the size of sampling domain. 

\subsection{Conditioning by Kriging}
In practice, hydraulic conductivity and transmissivity measurements are available from a few boreholes. However, the unconditional random field is really unrelated to these measurement data. One way to take into account the knowledge of the value of hydraulic parameter at the data locations is to use the spatial interpolation such as the kriging method \citep{delhomme:1979}. Although the kriging honors the actually observed value, it yields less varying estimators because the kriging coefficients are computed by the minimum variance of error. Therefore, kriged values are usually less dispersed than the actual values of the transmissivity on groundwater flow \citep{delhomme:1979}.  The best solution is a compromise between the unconditional simulation and the kriging method, which is called the conditional simulation. The conditional random field (CRF) is consistent with the available data at observation locations and has the same covariance as the realistic phenomenon.  

Consider a stationary Gaussian random field $Z(\mathbf{x})$ generated on a sampling grid $\Omega_1 = \{\mathbf{x}_1,\ldots,\mathbf{x}_p\}$ and known on the observation grid $\Omega_2 = \{\mathbf{x}_{p+1},\ldots,\mathbf{x}_n\}$. Now suppose that we have an unconditional random field $Z_S(\mathbf{x})$ independent of $Z(\mathbf{x})$ with the same covariance as $Z(\mathbf{x})$. Then unconditional random fields can be conditioned by the following transformation \citep{gotway:1994, delhomme:1979}: 
\begin{equation} \label{cond_transform}
Z_C(\textbf{x}) = \widehat{Z}(\textbf{x}) + \left(Z_S(\textbf{x})-\widehat{Z}_S(\textbf{x})\right),
\end{equation}
where $\widehat{Z}(\textbf{x})$ is the kriging estimator (see, e.g. \citep{delhomme:1979}) of $Z$ based on the measurements at locations $\textbf{x}_{p+1},\ldots,\textbf{x}_n$ and $\widehat{Z}_S(\textbf{x})$ is the kriging estimator of $Z_S$ using $Z_s(\textbf{x}_{p+1}),\ldots, Z_s(\textbf{x}_n)$. Since kriging is an exact interpolator, $\widehat{Z}(\mathbf{x}_i) = Z(\mathbf{x}_i)$ and $\widehat{Z}_S(\mathbf{x}_i) = Z_S(\mathbf{x}_i)$, so that $Z_C(\mathbf{x}_i) = Z(\mathbf{x}_i)$ if $\mathbf{x}_i \in \Omega_2$. Therefore, this modification of unconditional random field reproduces the measurement values.

Now if $\mathbf{Z}_{S_1}$ and $\mathbf{Z}_{S_2}$ are vectors of samples of $Z_S$ on the grid $\Omega_1$ and $\Omega_2$ respectively, the the random vector 
\begin{equation} \label{eq:ucond_vec}
\mathbf{Z} = \left( 
\begin{array}{c}
\mathbf{Z}_{S_1} \\
\mathbf{Z}_{S_2}
\end{array} 
\right),
\end{equation}
has the mean
\begin{equation} \label{eq:ucond_mean}
\mu = \left(
\begin{array}{c}
\mu_{1}  \\
\mu_{2}
\end{array}
\right),
\end{equation} 
and covariance 
\begin{equation} \label{eq:ucond_cov}
R = \left( 
\begin{array}{cc}
R_{11} & R_{12} \\
R_{21} & R_{22}
\end{array} 
\right),
\end{equation}
where the $R_{ij}$ are covariance matrices between two random vectors $\mathbf{Z}_{S_i}$ and $\mathbf{Z}_{S_j}$. 

Assuming $\mathbf{z}_2$ is an observation vector collected on $\Omega_2$, the random vector $\mathbf{Z}_{S_1}$ given $\mathbf{Z}_{S_2} = \mathbf{z}_2$, which is a vector of values of $Z_C(\mathbf{x})$ in (\ref{cond_transform}) with $\mathbf{x} \in \Omega_1$, can be written as 
\begin{equation}\label{eq:cond_rv}
\mathbf{Z}_C:= \varphi \left(\mathbf{Z}_{S_1}|\mathbf{Z}_{S_2}=\mathbf{z}_2\right) := P\mathbf{z}_2 + \left(\mathbf{Z}_{S_1} - P\mathbf{Z}_{S_2}\right),
\end{equation}
where $P =  R_{12}R_{22}^{-1}$ is a matrix of simple kriging weight. The difference term in parentheses on the -hand side of (\ref{eq:cond_rv}) can be regarded as a random vector of error caused by the smoothing. Hence, the conditional random vector $\mathbf{Z}_C$ corrects the kriged estimator obtained using data on $\Omega_2$ by adding the simulated estimation error. Note that the distribution of $\mathbf{Z}_C$ in (\ref{eq:cond_rv}) is again normal \citep{anderson:2003} with the mean
\begin{equation}\label{eq:cond_mean}
\tilde{\mu} = \mu_{1} + P(\mathbf{z}_2-\mu_{2}),
\end{equation}
and covariance
\begin{equation}\label{eq:cond_cov}
\tilde{R} = R_{11} - R_{12}R_{22}^{-1}R_{21}.
\end{equation}

In \citet{zang:2004} and \citet{chen:2008}, the KL expansion of the covariance function in (\ref{cond_transform}) was used to generate conditional log hydraulic conductivity field. However, the corresponding conditional covariance function is not spatially stationary. As a result, eigenvalues and eigenfunctions of the conditional covariance function usually need to be found numerically, and the computational cost is relatively high, especially for not separable covariance function used in (\ref{eq:exp_cov}). 
 
As mentioned in the previous section, the circulant embedding algorithm only works on rectangular grid. However the observation grid $\Omega_2$, in general, is not regular, so the entire grid $\Omega_1 \cup \Omega_2$ is not rectangular any more. \citet{dietrich:1996} resolved this problem via a new algorithm which enables us to generate unconditional random vectors from a rectangular sampling grid $\Omega_1$ and irregular observation grid $\Omega_2$ assuming $\Omega_1 \cap \Omega_2 = \emptyset$. Then each of unconditional random vectors is conditioned by kriging. Here we use this algorithm as a conditional random field generator.

\subsection{Multilevel Monte Carlo Method}
In this section, we briefly review the basic idea of the multilevel Monte Carlo (MLMC) technique. Suppose we are interested in finding the expected value of a linear of nonlinear functional $Q_M = \mathcal{Q}(\mathbf{Z}_M)$ of either conditional or unconditional random vector $\mathbf{Z}_M \in \mathrm{R}^M$. For example, a function $\mathcal{Q}(\mathbf{Z}_M)$ can be the travel time of a radioactive particle departing from the repository to the human environment as considered in this paper.  

Consider Monte Carlo simulations with different degrees of freedoms $M_\ell, \ell = 0,...,L$ such that  
\begin{equation}
M_\ell = s M_{\ell-1}, \qquad \mbox{for all } \ell = 1,...,L,
\end{equation}
where $s$ is a positive integer.

The main idea behind the MLMC technique is as follows. In contrast to the standard Monte Carlo (MC) approach in which all samples are generated on the finest level, samples on all grid level $M_\ell (\ell=0,\ldots,L)$ are taken into account in MLMC to estimate statistical moments of solution: 
\begin{eqnarray} \label{eq:telescopingsum}
\mathrm{E}[Q_{M_L}] &=& \mathrm{E}[Q_{M_0}] + \sum_{\ell=1}^L \mathrm{E}[Q_{M_\ell} - Q_{M_{\ell-1}}] \\
&=&\sum_{\ell=0}^L \mathrm{E}[Y_\ell],	\nonumber
\end{eqnarray}
where $Y_\ell := Q_{M_\ell} - Q_{M_{\ell-1}}$. For simplicity we have set $Y_0 := Q_{M_0}$. Each of these expectations is independently estimated in a way that the overall variance is minimized for a fixed computational cost.

Let $\widehat{Y}_\ell$ be an unbiased estimator for $\mathrm{E}[Y_\ell]$ using $N_\ell$ samples. The simplest estimator is the the standard MC estimator, which for $\ell > 0$,
\begin{equation}\label{eq:mc_est}
\widehat{Y}^{\mathrm{MC}}_{\ell,N_\ell} := \frac{1}{N_\ell} \sum_{i = 1}^{N_\ell} \left( Q_{M_\ell}^{(i)} - Q_{M_{\ell-1}}^{(i)}\right).
\end{equation}
It is important to note that the quantity $Q_{M_\ell}^{(i)} - Q_{M_{\ell-1}}^{(i)}$ in (\ref{eq:mc_est}) is computed from two discrete approximations with different grid size but the same random sample $\omega^{(i)}$. Then the variance of this estimator is $\mathrm{V}[\widehat{Y}_\ell] = \mathrm{V}[Y_\ell]/N_\ell$. 

The multilevel estimator is simply defined as
\begin{equation} \label{eq:mlmc_est}
\widehat{Q}^{\mathrm{ML}}_M := \sum_{\ell = 0}^{L} \widehat{Y}_\ell.
\end{equation}
Since all the expectations $\mathrm{E}[Y_\ell]$ are estimated independently, the variance of this multilevel estimator is
\begin{equation} \label{eq:var_multi}
\mathrm{V}[\widehat{Q}^{\mathrm{ML}}_M] = \sum_{\ell = 0}^{L}\frac{\mathrm{V}[Y_\ell]}{N_\ell}
\end{equation}

The computational cost of the multilevel Monte Carlo estimator is 
\begin{equation}\label{eq:mlmc_cost}
\mathcal{C}(\widehat{Q}^{\mathrm{ML}}_M) = \sum_{\ell = 0}^{L}N_\ell \mathcal{C}_\ell
\end{equation}
where $\mathcal{C}_\ell:=\mathcal{C}(Y_\ell^{(i)})$ represents the cost of a single sample of $Y_\ell$. If we treat the $N_\ell$ as continuous variable, then the variance is minimised for a fixed computational cost by choosing 
\begin{equation}
N_\ell \propto \sqrt{\mathrm{V}[Y_\ell]/\mathcal{C}_\ell}.
\end{equation}

In order to quantify the accuracy of the approximations, we consider the mean square error $\mathrm{MSE}(\widehat{Q},Q)$ of the estimator $\widehat{Q}$ as an estimator of $Q$
\begin{equation}
\mathrm{MSE}(\widehat{Q},Q) = \mathrm{V}[\widehat{Q}] + (\mathrm{E}[\widehat{Q}]-Q)^2.
\end{equation}
Then the mean square error of MLMC estimator in (\ref{eq:mlmc_est}) is
\begin{equation}\label{eq:mse_mlmc}
\mathrm{MSE}\left(\widehat{Q}^{\mathrm{ML}}_M, Q\right) = \sum_{\ell = 0}^{L}\frac{\mathrm{V}[Y_\ell]}{N_\ell} + \left( \mathrm{E}[Q_M - Q]\right)^2.
\end{equation}
To have the $\mathrm{MSE}(\widehat{Q}^{\mathrm{ML}}_M,Q)^{1/2} $ at the tolerance level $\epsilon$, we evenly distribute $\epsilon^2$ between both of terms on the right hand side of (\ref{eq:mse_mlmc}). It is obvious that, if $Q_M$ converges to Q in mean square, then $\mathrm{V}[Y_\ell] = \mathrm{V}[Q_{M_\ell} - Q_{M_{\ell-1}}]\rightarrow 0$   as $\ell \rightarrow \infty$, and so fewer samples required on finer levels to estimate $\mathrm{E}[Y_\ell]$. Furthermore, the coarsest level $\ell = 0$ can be kept fixed for all $\epsilon$, and so the cost per sample on level $\ell = 0$ does not grow as $\epsilon \rightarrow 0$. Therefore, the cost of MLMC estimator is cheaper than that of MC estimator in achieving the same tolerance level. 

\citet{cliffe:2011} make above analysis more precise (see their Theorem 1). In short, if there exist positive constants $\alpha,\beta,\gamma, C_\alpha, C_\beta, C_\gamma >0$ such that $\alpha \geq \frac{1}{2}\min(\beta,\gamma)$ and 
\begin{equation}
\begin{array}{ll}
(\mathit{i})&|\mathrm{E}[Q_{M_\ell}-Q]| \leq C_\alpha M_\ell^{-\alpha},\\
(\mathit{ii})&\mathrm{V}[Y_\ell] \leq C_\beta M_\ell^{-\beta},\\
(\mathit{iii})&\mathcal{C}_\ell \leq C_\gamma M_\ell^{-\gamma},
\end{array}
\end{equation}
then there exist a positive constant $C^{\mathrm{ML}}$, a value $L$ and a sequence $\{N_\ell\}_{\ell=0}^{L}$ such that $\mathrm{MSE}(\widehat{Q}_M^{\mathrm{\mathrm{ML}}},Q) < \epsilon^2$ for any $\epsilon < e^{-1}$, and 
\begin{equation} \label{eq:mlmc_asymp_cost}
\mathcal{C}_\epsilon(\widehat{Q}_M^{\mathrm{\mathrm{ML}}}) = \left\{ 
\begin{array}{ll}
         C^{\mathrm{ML}}\epsilon^{-2}, & \mbox{if $\beta < \gamma$},\\
		C^{\mathrm{ML}}\epsilon^{-2} (\log \epsilon)^2, & \mbox{if $\beta = \gamma$},\\
        C^{\mathrm{ML}}\epsilon^{-2-(\gamma-\beta)/\alpha}, & \mbox{if $\beta > \gamma$},\end{array}
\right.
\end{equation}
whereas 
\begin{equation}\label{eq:mc_asymp_cost}
\mathcal{C}_\epsilon(\widehat{Q}_M^{\mathrm{\mathrm{MC}}}) = C^{\mathrm{MC}} \epsilon^{-2-\gamma/\alpha}
\end{equation}
for some positive constant $C^{\mathrm{MC}}$.  

\subsection{Antithetic Variates}
The method of antithetic variates (AV) is one of most commonly used variance reduction methods in the standard MC technique due to its simplicity and ease of application \citep{dagpunar:2007}. In the context of multilevel Monte Carlo simulation, \citet{giles:2012} introduced antithetic multilevel Monte Carlo estimator for multidimensional SDEs driven by Brownian motions. However, their estimator is quite different from what used here: their antithetic estimator uses the average of antithetic pairs at a fine level $\ell$ instead of a single fine-level calculation, whereas we use antithetic pairs on both levels. For clarity, estimator used here is called the full antithetic estimator.    

The basic idea in antithetic variates is as follows. If $\widehat{Y}_\ell$  in (\ref{eq:mlmc_est}) is an unbiased estimator of $Y_\ell$, it is possible to obtain its antithetic pair, $\widehat{Y}_\ell^-$, having the same mean as $\widehat{Y}_\ell$ and a negative correlation with $\widehat{Y}_\ell$. Then the antithetic variates estimate
\begin{equation}\label{eq:av_est}
\widehat{Y}_\ell^{\mathrm{AV}} = \frac{\widehat{Y}_\ell+\widehat{Y}_\ell^-}{2} ,
\end{equation}
is again an unbiased estimator of $Y_\ell$. 

The variance of $\widehat{Y}_\ell^{\mathrm{AV}}$ is then written as 
\begin{eqnarray}
\mathrm{V}\left[\widehat{Y}_\ell^{\mathrm{AV}}\right] &=& \frac{\mathrm{V}[\widehat{Y}_\ell]+\mathrm{V}[\widehat{Y}_\ell^-] + 2 \mathrm{Cov}(\widehat{Y}_\ell, \widehat{Y}_\ell^-) }{4}\\
&=& \mathrm{V}[\widehat{Y}_\ell](1+\rho), \nonumber
\end{eqnarray}
where $\mathrm{Cov}(\widehat{Y}_\ell, \widehat{Y}_\ell^-)$ and $\rho$ are the covariance and the correlation coefficient between $\widehat{Y}_\ell$ and $\widehat{Y}_\ell^-$, respectively. Note that $\rho$ is negative when $\widehat{Y}_\ell^-$ and $\widehat{Y}_\ell$ are negatively correlated. It indicates that the value of $\mathrm{V}[\widehat{Y}_\ell^{\mathrm{AV}}]$ is smaller than $\mathrm{V}[\widehat{Y_\ell}]$ in standard MC. 

For the construction of antithetic counterpart of $\mathbf{Z}_C$, we first consider zero-mean random vectors $\mathbf{Z}^0_{S_i} = \mathbf{Z}_{S_i}-\mu_i$. Then the equation (\ref{eq:cond_rv}) can be rewritten as   
\begin{equation}\label{eq:cond_rv2}
\mathbf{Z}_C = \tilde{\mu} + (\mathbf{Z}^0_{S_1} - P\mathbf{Z}^0_{S_2}).
\end{equation}
This indicates that each of conditional random vectors is generated by adding the vector of error caused by kriging of the zero-mean random vector to the conditional mean in (\ref{eq:cond_mean}).
Then an antithetic counterpart of the random vector $\mathbf{Z}_C$ is simply computed by changing the sign of kriging errors:
\begin{equation}\label{eq:cond_rv_antithetic}
\mathbf{Z}_C^- = \tilde{\mu} - (\mathbf{Z}^0_{S_1} - P\mathbf{Z}^0_{S_2}).
\end{equation}
Note that a random vector $\mathbf{Z}_C^- $ has the same mean and covariance as in (\ref{eq:cond_mean}) and (\ref{eq:cond_cov}), respectively.

\section{Numerical Results}

\subsection{The Waste Isolation Pilot Plant Site Simulation}
\begin{figure} [h]
 \caption{Locations and $\log_{10}$ transmissivity data}
 \label{fig:observation}
 \noindent\includegraphics[width=20pc]{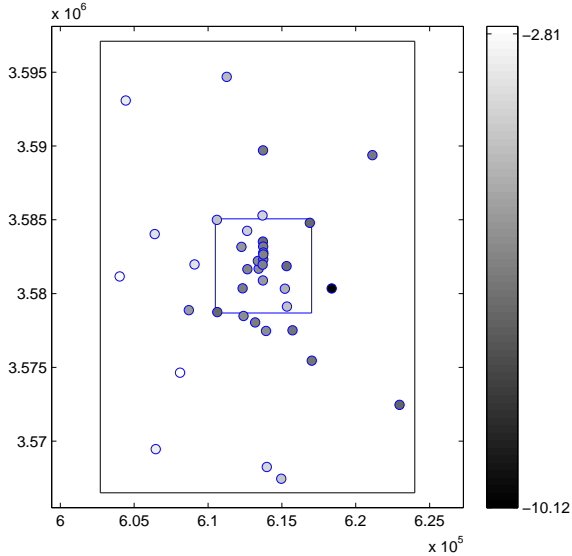}
 \end{figure}

The computational domain of the WIPP simulations performed in this paper is a rectangular region $D$ of Culebra Dolomite 21.5 km in east-west direction by 30.5 km in the north-south direction on the Universal Transverse Mercator (UTM) coordinates. The UTM system is an international location reference system that describes locations on a map. The WIPP site locates in the center of $D$. The $x$ and $y$ coordinates of 39 boreholes and the $\log_{10} T$ measurements \citep{cauffman:1990, nicola:2011} are presented in Table \ref{tab:wippdata}. Locations of these measurements are given in Figure \ref{fig:observation}, where the WIPP site boundary, $\partial \Gamma$, is shown by inner rectangle. These transmissivity observations will be used to predict groundwater pathline trajectory, velocity and travel time. 

\begin{table} 
\caption{WIPP Transmissivity Data \citep{cauffman:1990}}
\label{tab:wippdata}
\centering
\begin{tabular}{c c r l}
\hline
\hline
 X & Y & $\log_{10}T$  & Borehole \\
\hline
613,423&	3,581,684&	-6.03& H-1\\
612,660&	3,581,652&	-6.20& H-2\\
613,714&	3,580,892&	-5.61&  H-3\\
612,398&	3,578,484&	-6.00& H-4\\
616,888&	3,584,793&	-7.01& H-5\\
610,595&	3,584,991&	-4.45& H-6\\
608,106&	3,574,644&	-2.81& H-7\\
613,974&	3,568,252&	-3.90& H-9\\
622,967&	3,572,458&	-7.12 &H-10\\
615,341&	3,579,124&	-4.51& H-11\\
617,023&	3,575,452&	-6.71& H-12\\
612,341&	3,580,354&	-6.48& H-14\\
615,315&	3,581,859&	-6.88& H-15\\
613,369&	3,582,212&	-6.11& H-16\\
615,718&	3,577,513&	-6.64 &H-17\\
612,264&	3,583,166&	-5.78& H-18\\
615,203&	3,580,333&	-4.93& DOE-1\\
613,683&	3,585,294&	-4.02& DOE-2\\
609,084&	3,581,976&	-3.56& P-14\\
610,624&	3,578,747&	-7.04& P-15\\
613,926&	3,577,466&	-5.97& P-17\\
618,367&	3,580,350&	-10.12&P-18\\
613,710&	3,583,524&	-6.97& WIPP-12\\
612,644&	3,584,247&	-4.13& WIPP-13\\
613,735&	3,583,179&	-6.49& WIPP-18\\
613,739&	3,582,782&	-6.19 &WIPP-19\\
613,743&	3,582,319&	-6.57& WIPP-21\\
613,739&	3,582,653&	-6.40 &WIPP-22\\
606,385&	3,584,028&	-3.54& WIPP-25\\
604,014&	3,581,162&	-2.91& WIPP-26\\
604,426&	3,593,079&	-3.37& WIPP-27\\
611,266&	3,594,680&	-4.68 &WIPP-28\\
613,721&	3,589,701&	-6.60 &WIPP-30\\
613,696&	3,581,958&	-6.30& ERDA-9\\
613,191&	3,578,049&	-6.52& CB-1\\
614,953&	3,567,454&	-4.34& ENGLE\\
606,462&	3,569,459&	-3.26& USGS-1\\
608,702&	3,578,877&	-5.69& D-268\\
621,126&	3,589,381&	-6.55& AEC-7\\
\hline
\end{tabular}
\end{table} 
The primary quantity of interest in the safety assessment of geological disposal of the radioactive wastes is the travel time at which radionuclides released at the center of the WIPP site $\Gamma$ arrives to the WIPP site boundary $\partial \Gamma$. This is computed using the transport equation (\ref{eq:transport}) and the initial boundary condition (\ref{eq:transport_init}). We consider both the thickness of rock and the porosity to be constant and use $b = 8m$ and $\phi = 0.16$ as these are the values commonly used, see e.g. \citep{cauffman:1990,lavenue:1990}.  

\begin{figure} [h]
 \caption{Single sample calculation (using antithetic variates method).}
 \label{fig:algorithm_single_diagram}
 \noindent\includegraphics[width=20pc]{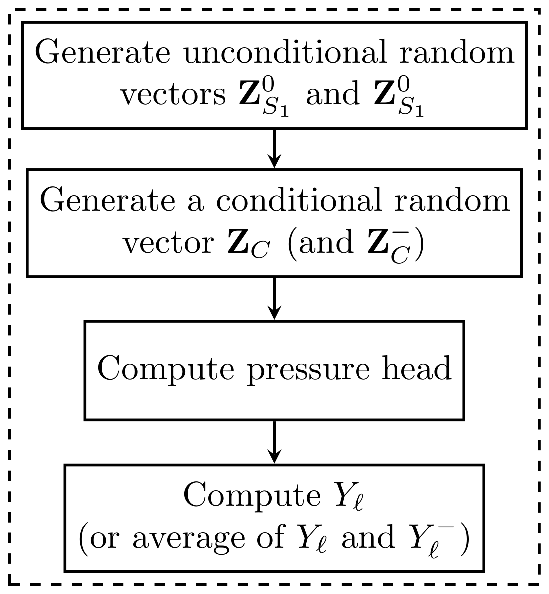}
 \end{figure}

\begin{figure} [h]
 \caption{MLMC algorithm for simulation of groundwater flow using conditional random fields.}
 \label{fig:algorithm_diagram}
 \noindent\includegraphics[width=20pc]{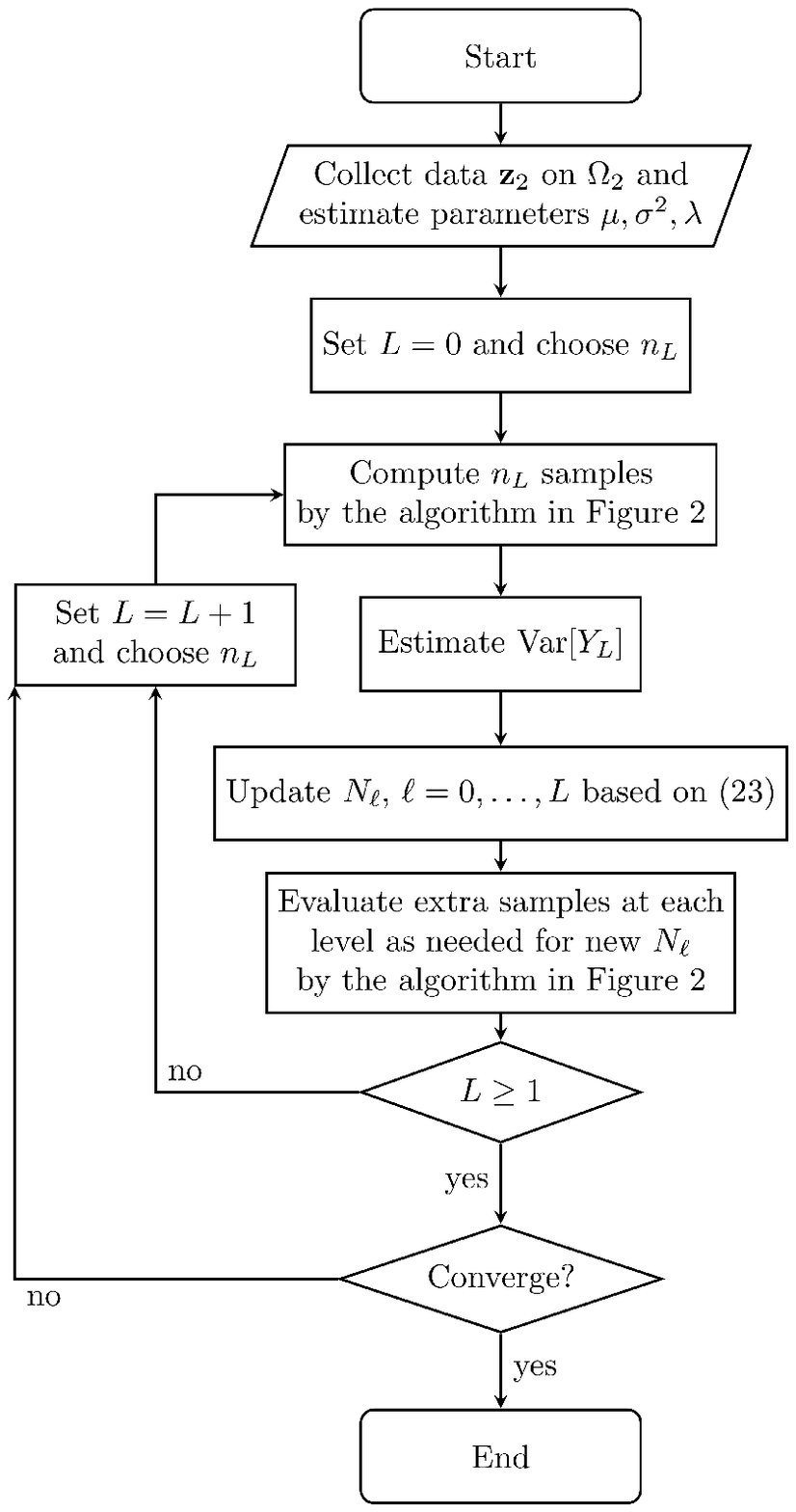}
 \end{figure}

Figures \ref{fig:algorithm_single_diagram} and \ref{fig:algorithm_diagram} present a flowchart illustrating the MLMC algorithm.   
Numerical solutions of (\ref{eq:groundwater}) are obtained with cell-centerd finite volume discretization of the groundwater flow problem \citep{cliffe:2011} using simulated conditional transmissivity fields generated by the circulant embedding methods recalled in Section \ref{sec:condrand}. We impose Dirichlet boundary conditions on the entire boundary $\partial D$. We use the following Gaussian function to compute the head values on the WIPP site boundaries (cf. (\ref{eq:dirichle})):
\begin{equation}
g_D(x,y) = a_0 \exp\left[ -\frac{\left(\left( \frac{x-x_0}{a_1}\right)^2 + \left( \frac{y-y_0}{a_2}\right)^2 \right)}{2} \right],
\end{equation}
where $a_0 = 1,134.61$, $a_1 = 73,559.35$, $a_2 = 73,559.35$, $x_0 = 611,011.89$ and $y_0 = 3,780,891.50$  \citep{cra2009}. We do not take into account the uncertainty arising from the approximation 	of this condition using the head measurements or from the head measurements.

We assume the log transmissivity vector has the same mean value in the entire domain $D$:
\begin{equation}\label{eq:const_mean}
\mathrm{E}[\mathbf{Z}_{S_i}(x)] =  \mu = \mbox{constant}, i = 1,2,
\end{equation}
where $x \in D$. In our numerical model, there are three unknown parameters, namely, the variance and correlation length in (\ref{eq:exp_cov}), and the constant mean in (\ref{eq:const_mean}). \citet{nicola:2011} derived the posterior distributions using Bayesian approach for the parameters, $\mu$, $\sigma^2$, and $\lambda$ of the log transmissivity field with a constant mean, and the mean values of the posterior distributions, which is listed in Table \ref{tab:parameters}, will be used as inputs to our numerical simulation.

\begin{table} [h]
\caption{Simulation parameters in (\ref{eq:exp_cov}) and (\ref{eq:const_mean})}
\label{tab:parameters}
\centering
\begin{tabular}{c | c }
\hline
 $\mu$ & -4.934\\
 $\sigma^2$ & 6.4791\\
 $\lambda$ & 12,390$m$\\
\hline
\end{tabular}
\end{table}

\begin{figure} [h]
\caption{The variance of $\log_{10}$ transmissivity field and locations of the measurements (+)}
\label{fig:var_samples}
\noindent\includegraphics[width=20pc]{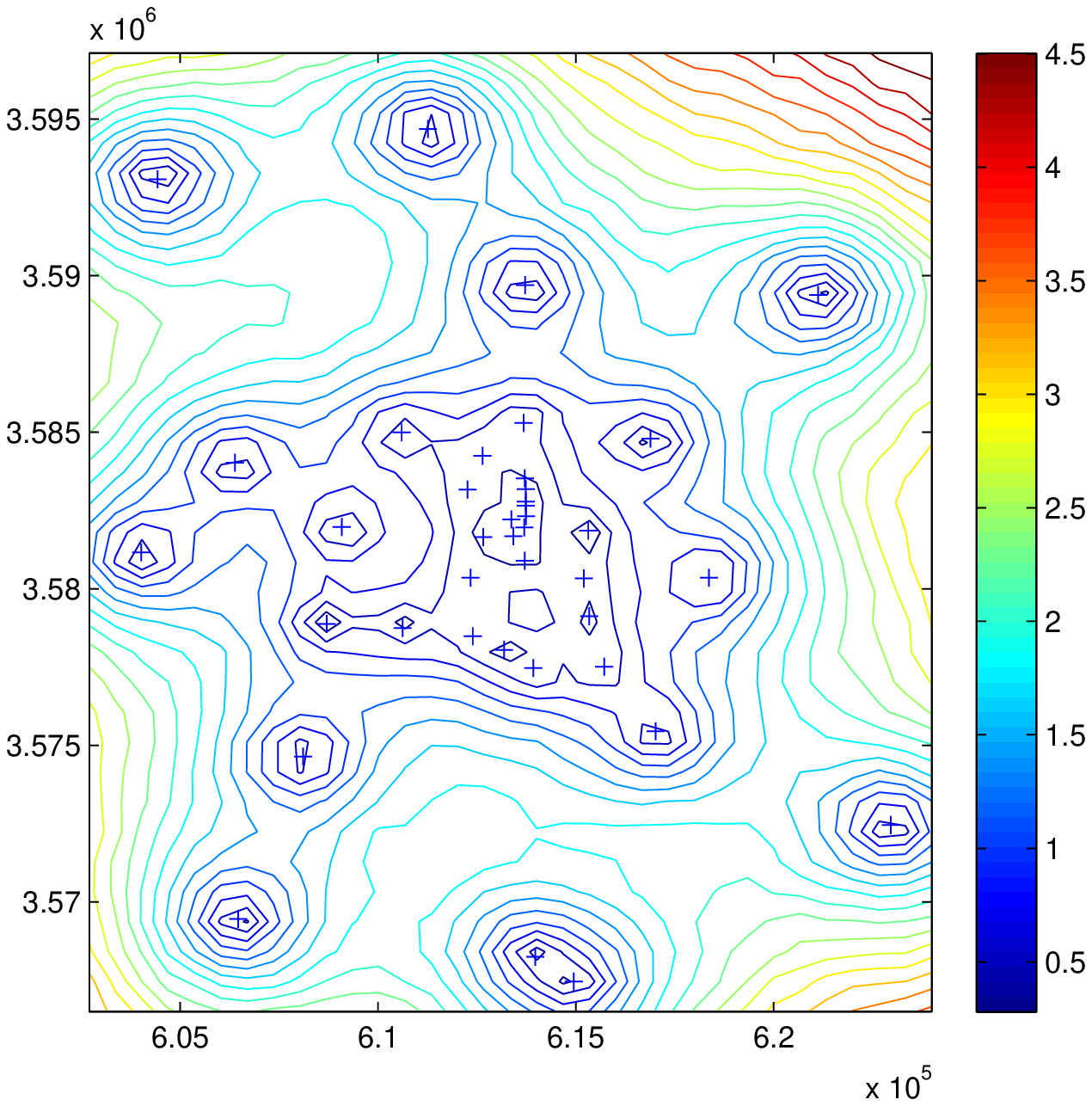}
\end{figure}
 
\begin{figure} [h]
\caption{The variance of pressure head using conditional random samples and locations of the measurements (+)}
\label{fig:var_head}
\noindent\includegraphics[width=20pc]{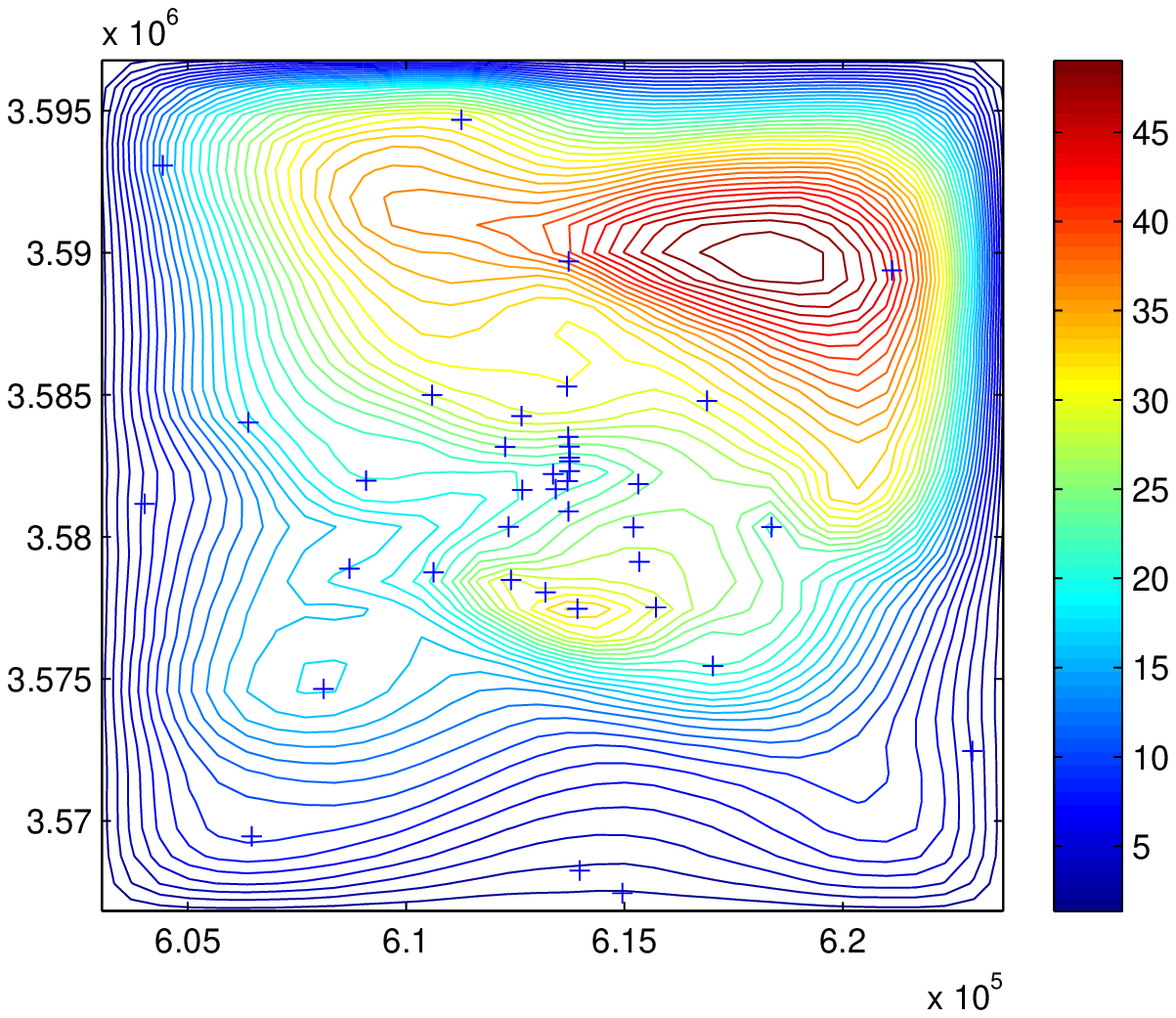}
\end{figure}

\subsection{Effect of Conditioning} \label{sec:effectofconditioining}
In this section, we use the WIPP site model to examine how the conditioning of random transmissivity field on observations in Table \ref{tab:wippdata} influences stochastic behavior of logarithm of the travel time. We generated 5000 realizations of unconditional random vectors $\mathbf{Z}_{S_1}$ and $\mathbf{Z}_{S_2}$ with covariance (\ref{eq:ucond_cov}) on each level $\ell$. Based on these realizations of the unconditional field, we build the same number of conditional random field  $\mathbf{Z}_C$ with the covariance in (\ref{eq:cond_cov}). 

\begin{figure} [h]
\caption{The variance of pressure head using unconditional random samples}
\label{fig:var_head_uncond}
\noindent\includegraphics[width=20pc]{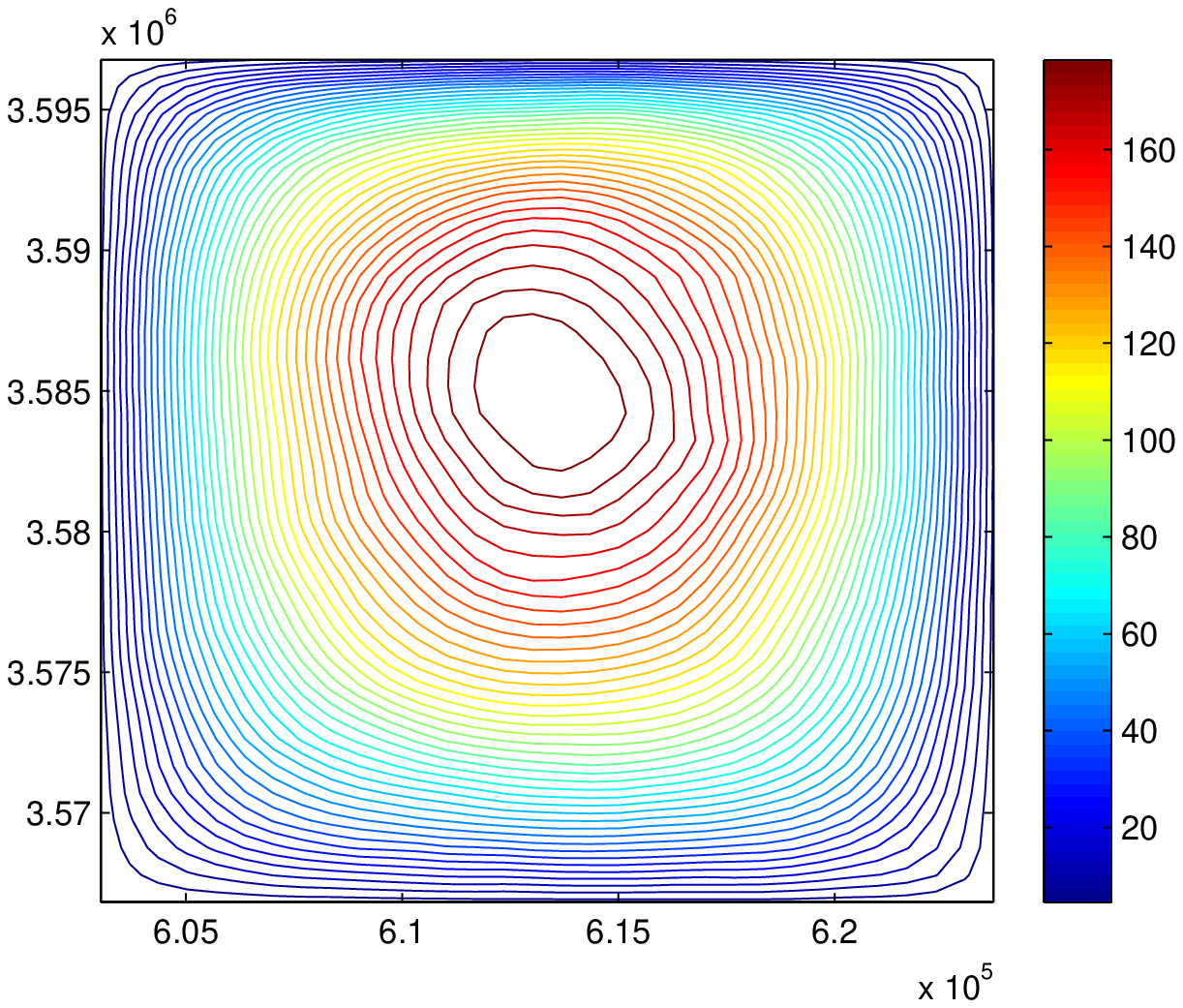}
\end{figure}

We first examine the effect of conditioning to transmissivity and head. Figure \ref{fig:var_samples} shows the variance of conditional log transmissivity field. The uncertainty of the log transmissivity is significantly diminished around the conditioning points. This variance reduction on the transmissivity random field then leads to a large reduction of overall predictive uncertainty of the head. Figures \ref{fig:var_head} and \ref{fig:var_head_uncond} compare the head variance derived from conditional and unconditional random fields. It is seen that the overall the head variance has been significantly reduced.    
 
\begin{figure}[h]
 \noindent\includegraphics[width=20pc]{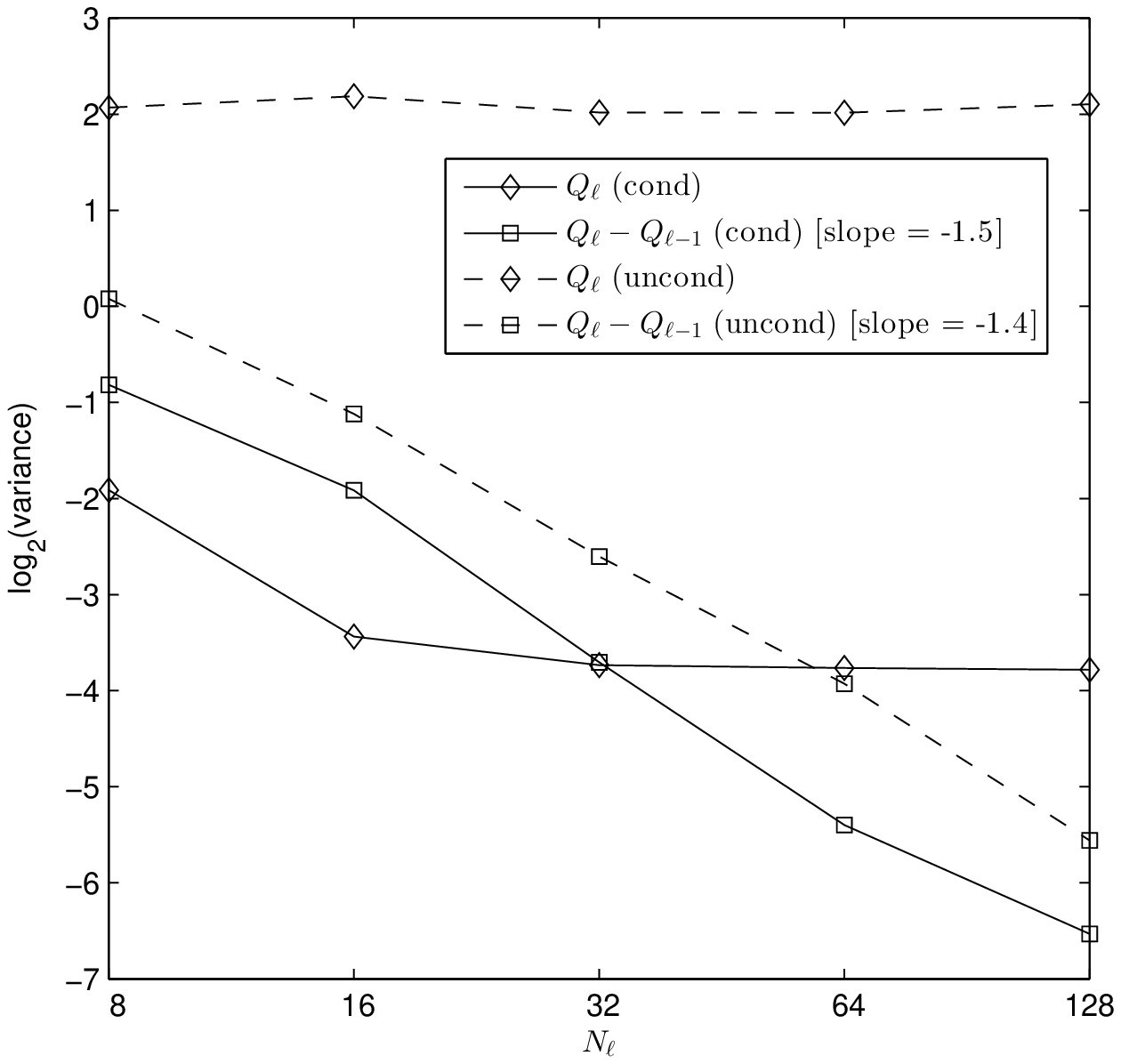}
 \caption{Variance plot of $\log_{10}$ travel time of a particle using the unconditional and conditional random fields.}
 \label{fig:var1}
 \end{figure}

Then the quantity of interest on level $\ell$, $Q_\ell$, and the difference between values of two consecutive levels, $Y_\ell = Q_\ell - Q_{\ell-1}$, are computed using 5000 realizations to see the effect of conditioning to the MLMC estimator. Figure \ref{fig:var1} compares the variances of logarithm of travel time. The variance profiles of $Y_\ell$ from both unconditional and conditional cases show the same behavior with a slope of almost -1.5, but the variance of $Y_\ell$ using the conditional fields is decreased by 50$\%$. The reduction in the variance of $Q_\ell$ by conditioning is even more significant. For the conditional case, the magnitude of $\mathrm{V}[Q_\ell]$ is 16 - 64 times smaller than that of the unconditional case. 

Consequently, a line for $\mathrm{V}[Q_\ell]$ is below that for $\mathrm{V}[Y_\ell]$ when the number of sampling points in each coordinate direction, $N_\ell$, is smaller than 32. When $\mathrm{V}[Q_\ell - Q_{\ell-1}] \geq \mathrm{V}[\mathrm{Q_\ell}]$, including level $\ell-1$, increases the cost of estimator. For this reason, we should exclude all levels coarser than $N_\ell = 32$ in the estimator. This indicates that the conditional MLMC simulations requires a finer mesh size on the coarsest level, which could result in the increase of complexity.

\subsection{MLMC Results}
\begin{figure}[h]
 \noindent\includegraphics[width=20pc]{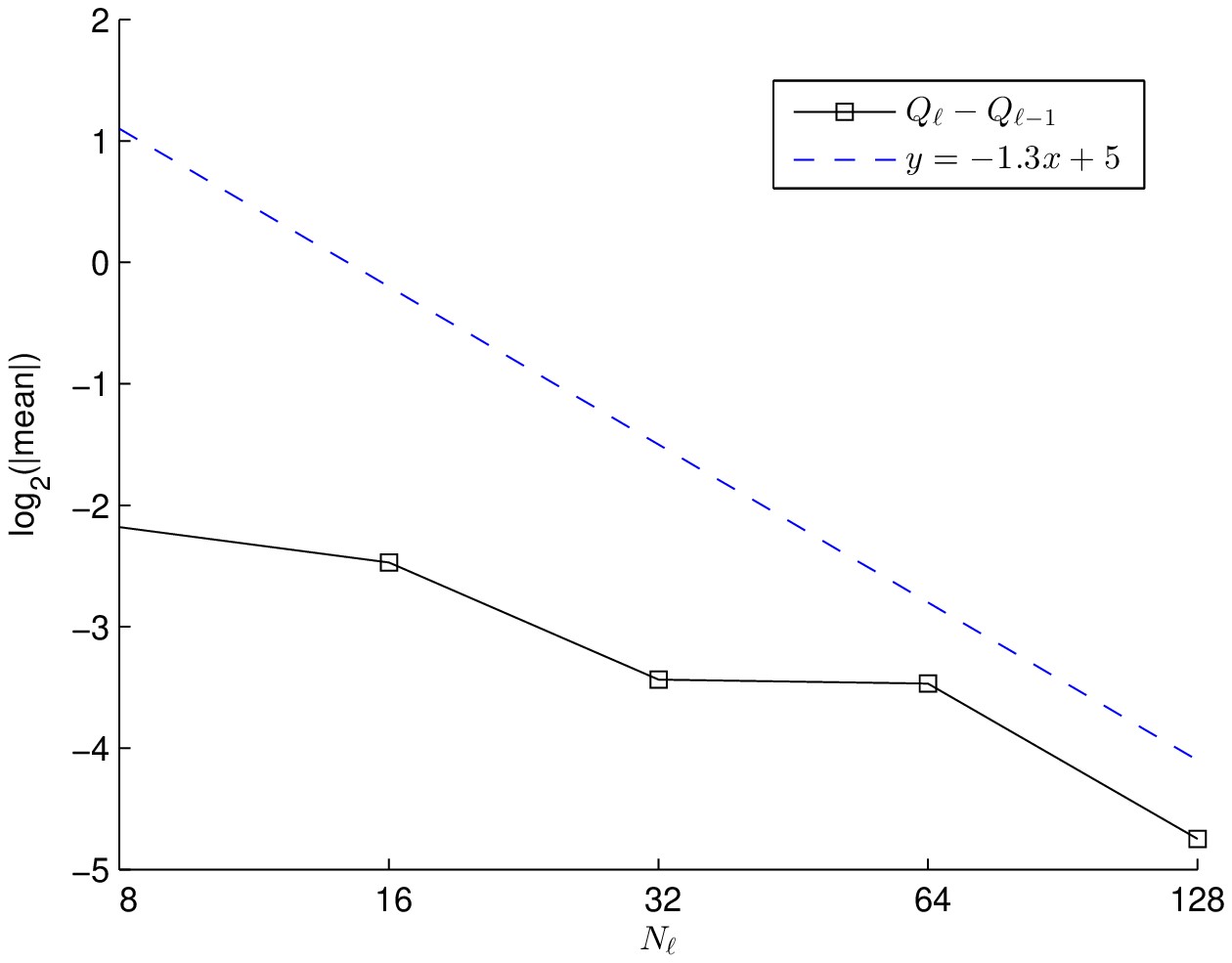}
 \caption{Mean plot of $\log_{10}$ travel time of a particle.}
 \label{fig:mean0}
 \end{figure}

In Figure \ref{fig:var1}, we have seen that the slope of line for $\mathrm{V}[Y_\ell]$ is roughly equal to -1.5, indicating that  $\mathrm{V}[Y_\ell] = c_1 h_\ell^{1.5} = c_1 M_\ell^{-0.75}$ for a positive constant $c_1$, or $\beta \approx 0.75$. Figure \ref{fig:mean0} shows the expected value of $Y_\ell$. As discussed in Section \ref{sec:effectofconditioining}, note that $N_\ell = 32$ is used as a problem size of the coarsest level, thus we consider the slope of the line for $\mathrm{E}[Y_\ell]$ between 64 and 128. Here we also present a straight line with the slope -1.3 for the comparison purpose. This implies again that $\mathrm{E}[Y_\ell] = c_2 h_\ell^{1.3} = c_2 M_\ell^{-0.65}$ for a positive constant $c_2$. As the iterative linear solver, we use an algebraic multigrid method, which is known as an optimal solver for the elliptic-type PDEs \citep{briggs:2000}. Hence, $\gamma = 1$. In Table \ref{tab:asymtotic_cost} we compare the predicted asymptotic order of $\mathcal{C}_\epsilon(\widehat{Q}^{\mathrm{ML}}_M)$ with $\mathcal{C}_\epsilon(\widehat{Q}^{\mathrm{MC}}_M)$ using (\ref{eq:mlmc_asymp_cost}) and (\ref{eq:mc_asymp_cost}). The results indicate that the MLMC estimator is asymptotically more efficient than the standard MC estimator when the observation are used in the simulation.

\begin{figure}[h]
 \noindent\includegraphics[width=20pc]{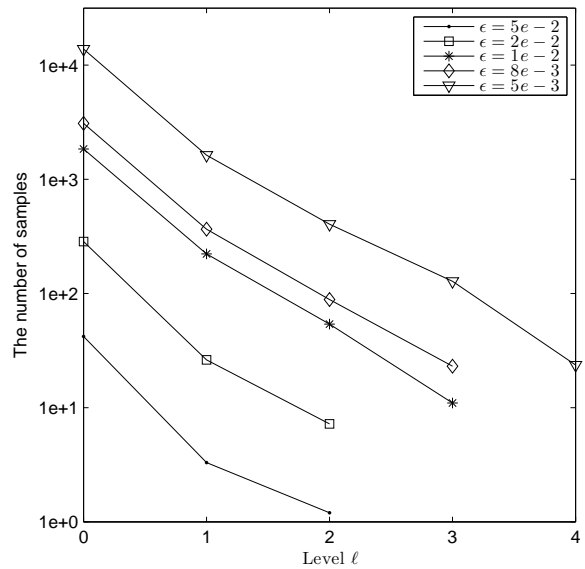}
 \caption{The number of samples used on each level}
 \label{fig:nsample}
 \end{figure}

\begin{figure}[h]
 \noindent\includegraphics[width=20pc]{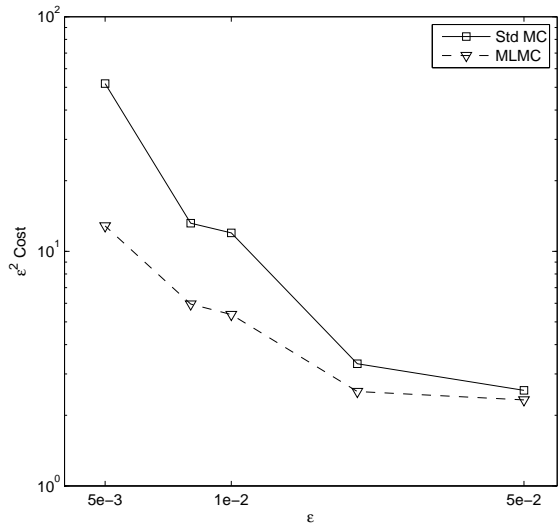}
 \caption{Comparison of the cost of standard MC with the cost of MLMC}
 \label{fig:cost_comparison}
 \end{figure}

Figure \ref{fig:nsample} shows the number of samples used on each level in the conditional MLMC simulations using an algorithm shown in Figures \ref{fig:algorithm_single_diagram} and \ref{fig:algorithm_diagram} and Figure \ref{fig:cost_comparison} presents comparison of the cost of the standard MC with the cost of MLMC. To quantify the cost of the algorithm in Figure \ref{fig:cost_comparison}, we assume that the number of operations to a single $Y_\ell^{(i)}$ is $\mathcal{C}_\ell = C^* M_\ell^\gamma$ for some constant $C^*$.  The results presented in Figure \ref{fig:cost_comparison} are the standardised costs, scaled by $1/C^*$. It is important to note that the conditioning is effective at reducing the variance of $Q_\ell$, resulting in the small constant in (\ref{eq:mc_asymp_cost}). Consequently, the cost-effectiveness is small as tolerance is large, e.g. $\epsilon = 5e-2$. However, the results show that the cost of the conditional MLMC glows more slowly than the cost of the conditional MC as $\epsilon \rightarrow 0$. 

 \begin{table} [h]
\caption{Predicted asymptotic order of cost to achieve a MSE of $\sqrt{\epsilon}$ in the case $\alpha = 0.65$, $\beta = 0.75$ and $\gamma = 1$ for the standard MC and MLMC estimators.}
\label{tab:asymtotic_cost}
\centering
\begin{tabular}{c | c }
\hline \\
$\mathcal{C}_\epsilon(\widehat{Q}^{\mathrm{MC}}_M)$ &$\mathcal{C}_\epsilon(\widehat{Q}^{\mathrm{ML}}_M)$\\ 
\\
\hline \\
 $\epsilon^{-3.53}$ & $\epsilon^{-2.38}$\\ \\
 \hline
\end{tabular}
\end{table}
 
\subsection{Effect of Antithetic Variates with Conditioning}
\begin{figure} [h]
 \caption{One pair of antithetic random field conditioned on given observations; both with covariance (\ref{eq:cond_cov}). The circles denote the location of observations. }
 \label{fig:antithetic_pair}
 \noindent\includegraphics[width=20pc]{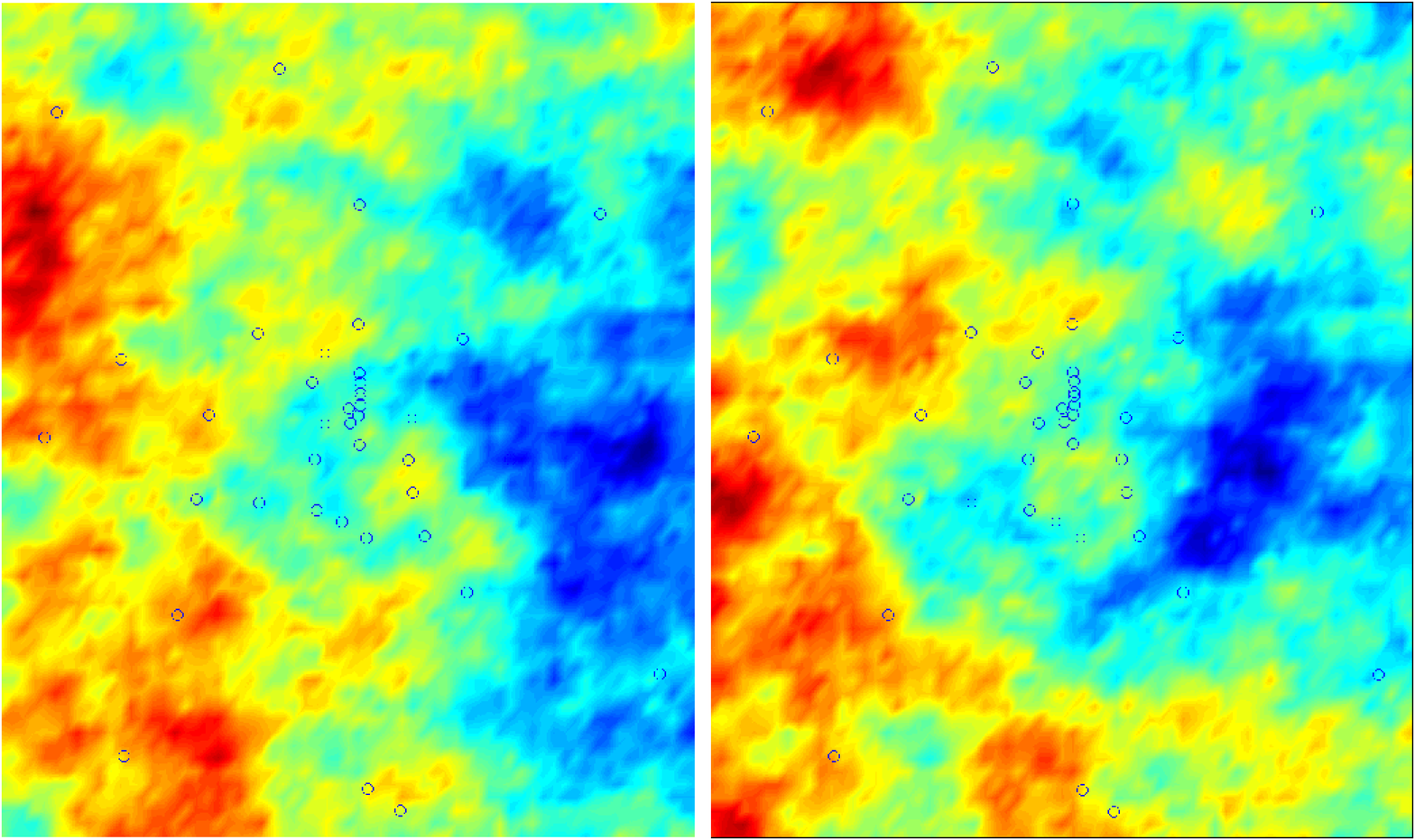}
 \end{figure}

The characteristic feature of antithetic variates is that one pair of unconditional antithetic random fields have exactly the reversed structure in comparison with its antithetic counterpart. In the conditional case shown in Figure \ref{fig:antithetic_pair} the influence of including observations is such, that in the local neighborhood of the measurements the variation is small, while in the rest of the domain, the structure is reversed.

\begin{figure}[h]
 \noindent\includegraphics[width=20pc]{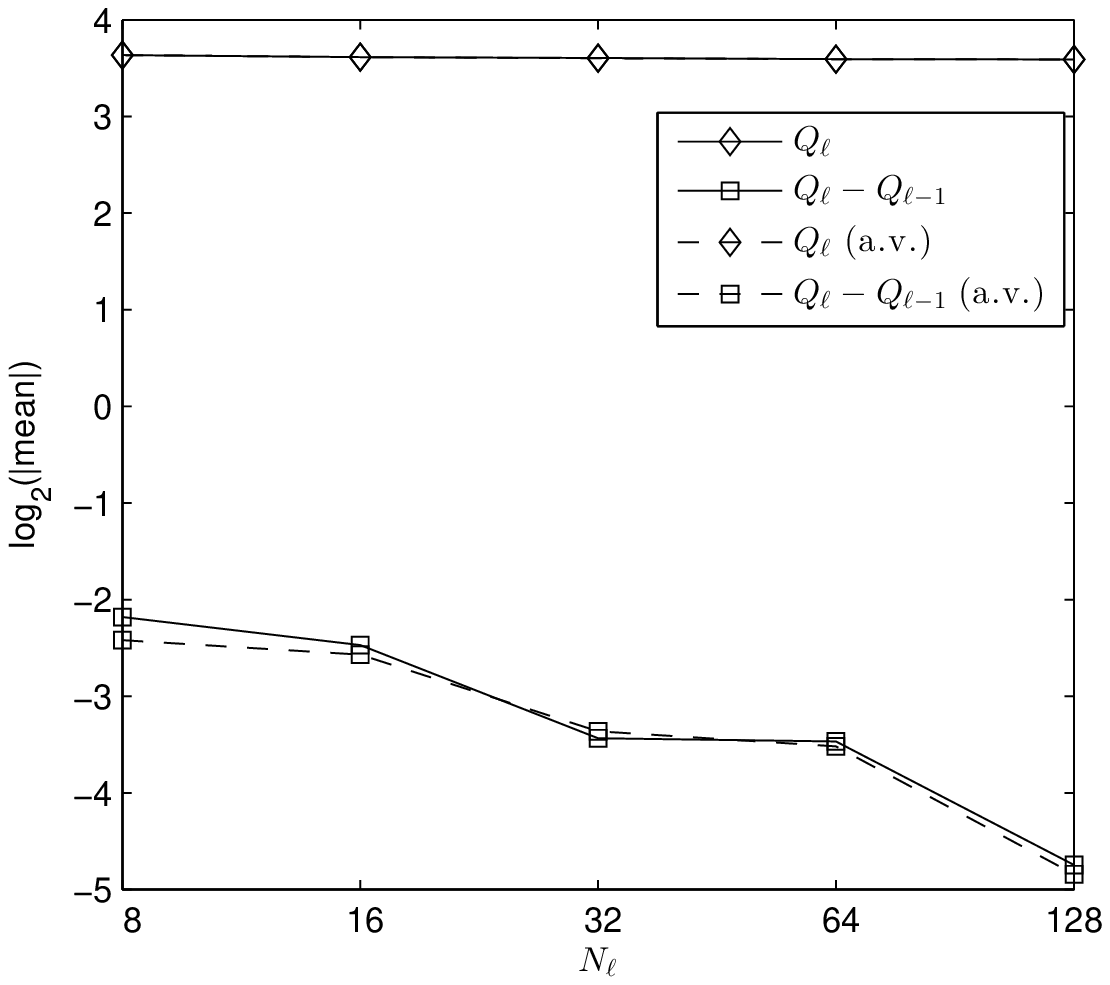}
 \caption{Mean plot of $\log_{10}$ travel time of a particle. a.v. stands for antithetic variates.}
 \label{fig:mean}
 \end{figure}

\begin{figure}[h]
 \noindent\includegraphics[width=20pc]{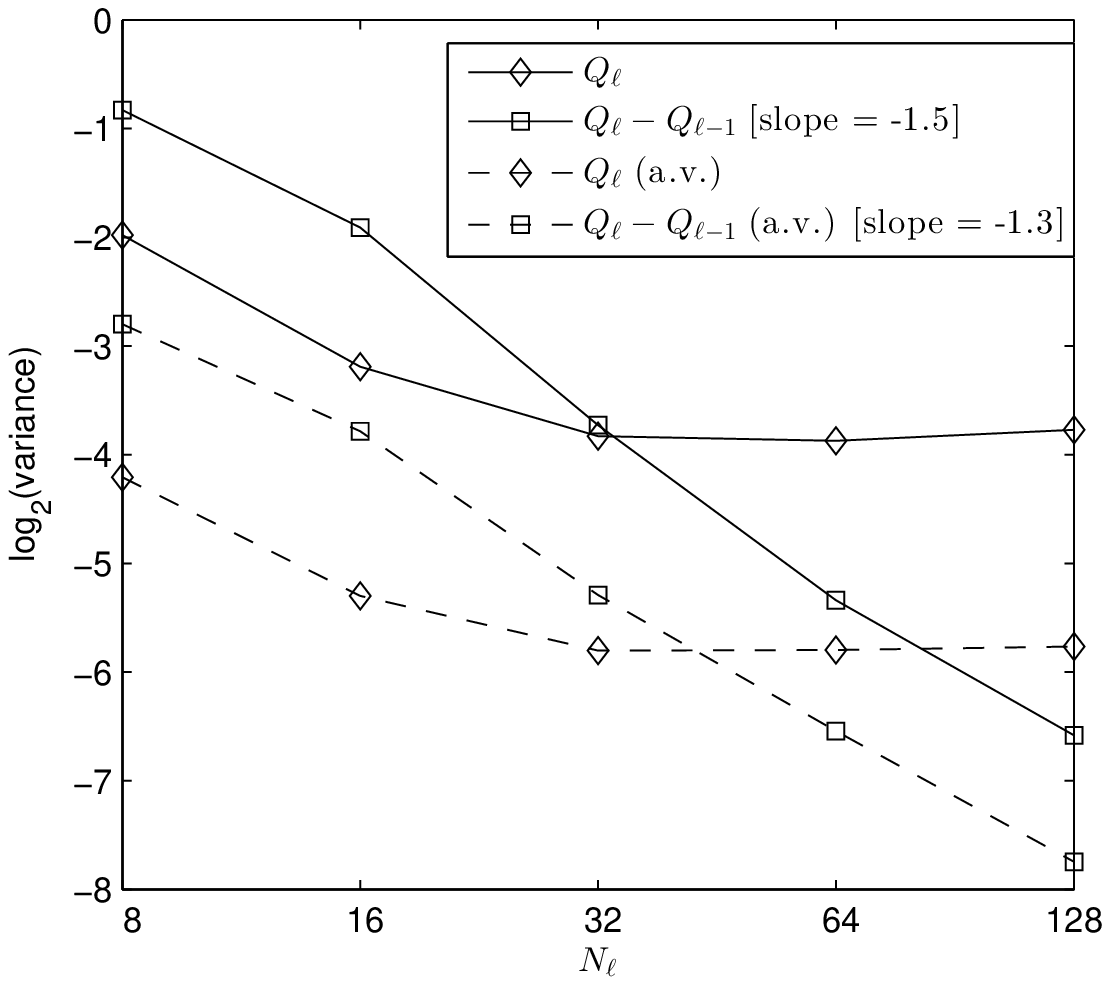}
 \caption{Variance plot of $\log_{10}$ travel time of a particle.}
 \label{fig:var}
 \end{figure}

Figures \ref{fig:mean} and \ref{fig:var} show the mean and variance plot of $Q_\ell$ and $Y_\ell$ with and without the antithetic variates method. Since the full antithetic variates estimator is unbiased, two results shows identical expected values. Using antithetic variates (dashed line), we see that variances of $Q_\ell$ and $Y_\ell$ are reduced by a factor of 4 and 2, respectively. Nevertheless, 	the use of antithetic variates method does not improve the slope of a line for $\mathrm{V}[Y_\ell]$, which indicates that the ratio of saving to cost by using the antithetic variates methods is constant as $\epsilon \rightarrow 0$. 
 
 We examine efficiency of antithetic variates in the MLMC simulations by measuring the CPU time of each of given simulations which were performed on a server with Intel Xeon E5-2450 8-core processors and 96Gb RAM running under Linux. The program is written in C$/$C++ and is compiled by GCC$/$G++ compilers. In Table \ref{tab:cputime}, we observe almost 45$\%$ reductions in computational costs by using the antithetic variates over conditional MLMC simulations. As we expected, the rate of saving is nearly the same in all cases.  
\begin{table} [h]
\caption{CPU time comparison of MLMC with a combination of MLMC and A.V. in seconds.}
\label{tab:cputime}
\centering
\begin{tabular}{c || c | c | c}
\hline
& & & \\
 $\epsilon$  & MLMC & MLMC + A.V.  & ratio\\
& & & \\
\hline 
  1e-2 &  1.95e+03 & 1.36e+03 & 0.697 \\
  8e-3 &  3.19e+03 & 2.09e+03 & 0.655\\
  5e-3 &  2.49e+04 & 1.56e+04 & 0.626\\
\hline
\end{tabular}
\end{table}

\section{Conclusions and Future Work}
The multilevel Monte Carlo simulation technique for simulation of groundwater flow using conditional transmissivity random field has been considered in this paper. We observed that the conditioning is more effective at reducing variance of the estimator on the coarsest level. The conditional MLMC simulation, thus, requires to start from a finer coarsest grid, which could increase the overall complexity of this method. However, numerical results show the advantage of using the MLMC estimator over a standard MC estimator for the groundwater flow simulation, which involves direct measurements of porous medium properties. We also used the antithetic variates for further variance reduction of the MLMC estimator. 

In this paper, the main goal was to examine the influence by conditioning on the MLMC simulations. Since estimation of cumulative distribution function of the travel time is also of great interest in groundwater flow research, hence the potential for the future work is to employ an cumulative distribution function estimation in the MLMC simulations.


%
\section*{Acknowledgements}
The authors are very grateful to Prof. Michael Tretyakov at the University of Nottingham, United Kingdom, for his valuable comments on the manuscript. This research was funded by the Engineering and Physical Sciences Research Council under the research grant EP/H051589/1.


\begin{thebibliography}{25}
\providecommand{\natexlab}[1]{#1}
\expandafter\ifx\csname urlstyle\endcsname\relax
  \providecommand{\doi}[1]{doi:\discretionary{}{}{}#1}\else
  \providecommand{\doi}{doi:\discretionary{}{}{}\begingroup
  \urlstyle{rm}\Url}\fi

\bibitem[{\textit{Anderson}(2003)}]{anderson:2003}
Anderson, T. (2003), \textit{An {I}ntroduction to {M}ultivariate {S}tatistical
  {A}nalysis}, 3 ed., John Wiley \& Sons, Inc., New Jersey, USA.

\bibitem[{\textit{Barth et~al.}(2011)\textit{Barth, Schwab, and
  Zollinger.}}]{barth:2011}
Barth, A., C.~Schwab, and N.~Zollinger. (2011), Multi-level {M}onte {C}arlo
  finite element method for elliptic {PDE}s with stochastic coefficients.,
  \textit{Numerische Mathematik}, \textit{119}, 123--161.

\bibitem[{\textit{Briggs et~al.}(2000)\textit{Briggs, Henson, and
  McCormick}}]{briggs:2000}
Briggs, W., V.~E. Henson, and S.~McCormick (2000), \textit{A {M}ultigrid
  {T}utorial}, 2 ed., SIAM, Philadelphia.

\bibitem[{\textit{Cauffman et~al.}(1990)\textit{Cauffman, LaVenue, and
  McCord}}]{cauffman:1990}
Cauffman, T., A.~LaVenue, and J.~McCord (1990), Ground-water flow modeling of
  the culebra dolomite. volume \textrm{II}: Data base., \textit{Tech. Rep.
  SAND89-7068/2}, Sandia National Laboratories.

\bibitem[{\textit{Chen et~al.}(2008)\textit{Chen, Lu, and
  Zyvolosk}}]{chen:2008}
Chen, M., Z.~Lu, and G.~A. Zyvolosk (2008), Conditional simulation of water-oil
  flow in heterogeneous porous media, \textit{Stoch. Environ. Res. Risk
  Asses.}, \textit{22}(4), 587--596.

\bibitem[{\textit{Cliffe et~al.}(2011)\textit{Cliffe, Giles, Scheichl, and
  Teckentrup}}]{cliffe:2011}
Cliffe, K.~A., M.~Giles, R.~Scheichl, and A.~L. Teckentrup (2011), Multilevel
  {M}onte {C}arlo methods and applications to elliptic {PDE}swith random
  coefficients., \textit{Comput. Visual. Sci.}, \textit{14}(1), 3--15.

\bibitem[{\textit{Dagan}(1982)}]{dagan:1982}
Dagan, G. (1982), Stochastic modeling of groundwater flow by unconditional and
  conditional probabilities. 1. {C}onditional simulations and the direct
  problem, \textit{Water Resour. Res.}, \textit{18(4)}, Daga:1982.

\bibitem[{\textit{Dagpunar}(2007)}]{dagpunar:2007}
Dagpunar, J. (2007), \textit{Simulation and {M}onte {C}arlo with application in
  finance and MCMC}, John Wiley \& Sons Ltd, West Sussex, England.

\bibitem[{\textit{de~Marsily et~al.}(2005)\textit{de~Marsily, Delay, Goncalves,
  Renard, Teles, and Violette}}]{marsily:2005}
de~Marsily, G., F.~Delay, J.~Goncalves, P.~Renard, V.~Teles, and S.~Violette
  (2005), Dealing with spatial heterogeneity, \textit{Hydrogeol. J.},
  \textit{13}, 161--183.

\bibitem[{\textit{Delhomme}(1979)}]{delhomme:1979}
Delhomme, J. (1979), Spatial variability and uncertainty in groundwater flow
  parameters, a geostatistical approach, \textit{Water Resourc. Res.}, pp.
  269--280.

\bibitem[{\textit{Dietrich and Newsam}(1989)}]{dietrich:1989}
Dietrich, C.~R., and G.~Newsam (1989), A stability analysis of the
  geostatistical approach to aquifer identification, \textit{Stochastic Hydrol.
  Hydraul.}, \textit{4}(3), 293--316.

\bibitem[{\textit{Dietrich and Newsam}(1996)}]{dietrich:1996}
Dietrich, C.~R., and G.~Newsam (1996), A fast and exact method for
  multidimensional {G}aussian stochastic simulations: Extension to realizations
  conditioned on direct and indirect measurements, \textit{Water Resour. Res.},
  \textit{32(6)}, 1643--1652.

\bibitem[{\textit{Dietrich and Newsam}(1997)}]{dietrich:1997}
Dietrich, C.~R., and G.~Newsam (1997), Fast and exact simulation of stationary
  {G}aussian processes through circulant embedding of the covariance matrix,
  \textit{SIAM J. SCI. COMPUT.}, \textit{18}(4), 1088--1107.

\bibitem[{\textit{Ghanem and Spanos}(1991)}]{ghanem:1991}
Ghanem, R., and D.~Spanos (1991), \textit{Stochastic Finite eelement: a
  spectral approach.}, Springer, New York.

\bibitem[{\textit{Giles and Szpruch}(2012)}]{giles:2012}
Giles, M., and L.~Szpruch (2012), Antithetic mulilevel {M}onte {C}arlo
  estimation for multi-dimentional {SDE}s without {L}$\acute{\mbox{e}}$vy area
  simulation, \textit{arXiv:1202.6283}.

\bibitem[{\textit{Gotway}(1994)}]{gotway:1994}
Gotway, C.~A. (1994), The use of conditioinal simulation in nuclear-waste-site
  performance assessment, \textit{Technometrics}, \textit{36}(2), 129--141.

\bibitem[{\textit{Graham and McLaughlin}(1989)}]{graham:1989}
Graham, W., and D.~McLaughlin (1989), Stochastic analysis of nonstationary
  subsurface solute transport, 2. {C}onditioinal moments, \textit{Water Resour.
  Res.}, \textit{25}(11), 2331--2355.

\bibitem[{\textit{LaVenue et~al.}(1990)\textit{LaVenue, RamaRao, and
  Reeves}}]{lavenue:1990}
LaVenue, A.~M., B.~S. RamaRao, and M.~Reeves (1990), Ground-water flow modeling
  of the culebra dolomite at the waste isolation pilot plant (wipp) site : Vol.
  1. model calibration. vol. 2. data base, \textit{Tech. Rep. SAND89-7068/1,
  SAND-7068/2.}, Sandia National Laboratories.

\bibitem[{\textit{Le~Ma\^{i}tre and Kino}(2010)}]{maitre:2010}
Le~Ma\^{i}tre, O., and O.~Kino (2010), \textit{Spectral Methods for Uncertainty
  Quantification, With Applications to Fluid Dynamics}, vol. Berlin, Springer.

\bibitem[{\textit{Lu and Zhang}(2004)}]{lu:2004}
Lu, Z., and D.~Zhang (2004), Conditional simulations of flow in rarandom
  heterogeneous porous media using a {KL}-based moment-equation approach,
  \textit{Adv. Water Resour.}, \textit{27}, 859--874.

\bibitem[{\textit{Stone}(2011)}]{nicola:2011}
Stone, N. (2011), Gaussian process emulators for uncertainty analysis in
  groundwater flow, Ph.D. thesis, The University of Nottingham.

\bibitem[{\textit{Teckentrup et~al.}(2013)\textit{Teckentrup, Scheichl, Giles,
  and Ullmann.}}]{teckentrup:2013}
Teckentrup, A., R.~Scheichl, M.~Giles, and E.~Ullmann. (2013), Further analysis
  of multilevel monte carlo methods for elliptic pdes with random
  coefficients., \textit{Numerische Mathematik,}, p. 1–32.

\bibitem[{\textit{{U.S. D.O.E.}}(2009)}]{cra2009}
{U.S. D.O.E.} (2009), 2009 {WIPP} compliance recertification application.,
  http://www.wipp.energy.gov/Documents$\_$EPA.htm.

\bibitem[{\textit{Xiu}(2010)}]{xiu:2010}
Xiu, D. (2010), \textit{Numerical Methods for Stochastic Computations: A
  Spectral Method Approach}, Princeton University Press, Princeton.

\bibitem[{\textit{Zang and Lu}(2004)}]{zang:2004}
Zang, D., and Lu (2004), Evaluation of higher-order moments for saturated flow
  in rarandom heterogeneous media via {K}arhunen-{L}oeve decomposition,
  \textit{J. Comput. Phys.}, \textit{194}(2), 773--794.

\end{thebibliography}
\end{document}